# Luminescent ions in silica-based optical fibers

Bernard Dussardier*, Wilfried Blanc and Gérard Monnom

Laboratoire de Physique de la Matière Condensée, Université de Nice Sophia-Antipolis - CNRS, UMR 6622, Parc Valrose - F-06108 NICE CEDEX 2 – France

* Corresponding author: email: bernard.dussardier@unice.fr, tél: +33 492 076 748

**Abstract**

We present some of our research activities dedicated to doped silica-based optical fibers, aiming at understanding the spectral properties of luminescent ions, such as rare-earth and transition metal elements. The influence of the local environment on dopants is extensively studied: energy transfer mechanisms between rare-earth ions, control of the valence state of Chromium ions, effect of the local phonon energy on Thulium ions emission efficiency and broadening of Erbium ions emission induced by oxide nanoparticles. Knowledge of these effects is essential for photonics applications.

**Introduction**

During the last two decades, the development of sophisticated optical systems and devices based on fiber optics have benefited from the development of very performant optical fiber components. In particular, optical fibers doped with 'active' elements such as rare-earth (RE) ions have allowed the extremely fast development of optical telecommunications [i,ii], lasers [iii] industries and the development of temperature sensors [iv]. The most frequently used RE ions ($Nd^{3+}$, $Er^{3+}$, $Yb^{3+}$, $Tm^{3+}$) have applications in three main spectral windows: around 1, 1.5 and 2µm in fiber lasers and sensors based on absorption/fluorescence and around 1.5 µm for telecommunications and temperature sensors. RE-doped fibers are either doped with one element (e.g. $Er^{3+}$ in line amplifiers for long haul telecommunications) or two elements (e.g. $Yb^{3+}$ and $Er^{3+}$ in booster amplifiers or powerful 1.5 µm lasers). In the second case, the non-radiative energy transfer mechanism from donor to acceptor is implemented to benefit from the good pump absorption capacity of the donor (e.g. $Yb^{3+}$ around 0.98 µm) and from the good stimulated emission efficiency of the acceptor (e.g. $Er^{3+}$ around 1.5 µm). All the developed applications of amplifying optical fibers are the result of long and careful optimization of the material properties, particularly in terms of dopant incorporation in the glass matrix, transparency and quantum efficiency.

The exploited RE-doped fibers are made of a choice of glasses: silica is the most widely used, sometimes as the result of some compromises. Alternative glasses, including low Maximum Phonon Energy (MPE) ones, are also used because they provide better quantum efficiency or emission bandwidth to some RE ions particular optical transition. The icon example is the $Tm^{3+}$-Doped Fiber Amplifier (TDFA) for telecommunications in the S-band (1.48-1.53 µm) [v], for which low MPE glasses have been developed: oxides [vi,vii], fluorides [viii], chalcogenides [ix]… However, these glasses have some drawbacks not acceptable at a commercial point of view: high fabrication costs, low reliability, difficult connection to silica components and, in the case of fiber lasers, low optical damage threshold and resistance to heat. To our knowledge, silica glass is the only material able to meet most of applications requirements, and therefore the choice of vitreous silica for the active fiber material is of critical importance. However a pure silica TDFA would suffer strong non-radiative de-excitation (NRD) caused by multiphonon coupling from $Tm^{3+}$ ions to the matrix. Successful insulation of $Tm^{3+}$-ions from matrix vibrations by appropriate ion-site 'engineering' would allow the development of a practical silica-based TDFA.

Other dopants have recently been proposed to explore amplification over new wavelength ranges. Bi-doped glasses with optical gain [x] and fiber lasers operating around 1100-1200 nm have been developed [xi,xii], although the identification of the emitting center is still not clear, and optimization of the efficiency is not yet achieved. Transition metal (TM) ions of the Ti-Cu series would also have interesting applications as broad band amplifiers, super-fluorescent or tunable laser sources, because they have in principle ten-fold spectrally larger and stronger emission cross-sections than RE ions. However, important NRD strongly reduces the emission quantum efficiency in silica. Bi- and TM-doped fibers optical properties are extremely sensitive to the glass composition and/or structure to a very local scale. As for $Tm^{3+}$ ions, practical applications based on silica would be possible when the 'ion site engineering' will be performed in a systematic approach. This approach is proposed via 'encapsulation' of dopants inside glassy or crystalline nanoparticles (NP) embedded in the fiber glass, like reported for oxyfluoride fibers [xiii] and multicomponent silicate fibers [xiv]. In NP-doped-silica fibers, silica would act as support giving optical and mechanical properties to the fiber, whereas the dopant spectroscopic properties would be controlled by the NP nature. The NP density, mean diameter and diameter distribution must be optimized for transparency [xv].

In this context, our group has made contributions in various aspects introduced above. Our motivations are both fundamental and application oriented. First, the selected dopants act as probes of the local matrix environment, via their spectroscopic variations versus ligand field intensity, site structure, phonon energy, statistical proximity to other dopants,… The studies are always dedicated to problems or limitation in applications, such as for Erbium-Doped Fiber Amplifier (EDFA) and TDFA, or high temperature sensors. It is also important to use a commercially derived fabrication technique, here the Modified Chemical Vapor Deposition (MCVD), to assess the potential of active fiber components for further development.

The aim of this paper is reviewing our contributions to improving the spectroscopic properties of some RE and TM ions doped into silica. The article is organized as follows: Section 0 describes the MCVD fabrication method of preform and fiber samples, and the common characterization techniques used in all studies. Section 0  Chapitre: is devoted to the study of energy transfers in Erbium ion ($Er^{3+}$) and Ytterbium:Erbium ($Yb^{3+}$:$Er^{3+}$) heavily (co)doped fibers and the applications to fiber temperature sensors, whereas Section 0 summarizes our original investigations on Chromium ($Cr^{3+}$ and $Cr^{4+}$) in silica-based fibers. In Section 0 Chapitre:, we report on the spectroscopic investigations of Thulium- ($Tm^{3+}$) doped fibers versus the material composition, including phonon interactions and non-radiative relaxations. In section 0 Chapitre: are reported our recent discoveries in RE-doped dielectric nanoparticles, grown by phase separation.

**Experimental**

**Preforms and fibers fabrication**

All the fibers investigated in this article were drawn from preforms prepared by the Modified Chemical Vapor Deposition (MCVD) technique [xvi] at Laboratoire de Physique de la Matière Condensée (Nice). In this process, chemicals (such as $O_2$, $SiCl_4$) are mixed inside a glass tube that is rotating on a lathe. Due to the flame of a burner moving along the tube, they react and extremely fine silica particles are deposited on the inner side of the tube. These soot are transformed into a glass layer (thickness is about few μm) when the burner is passing over. The cladding layers are deposited inside the substrate tube, followed by the core layers. Germanium and Phosphorus can be incorporated directly through the MCVD process. They are added to raise the refractive index. Moreover, this last element is also added as a melting agent, decreasing the melting temperature of the glass. All the other elements (rare-earths, transition metals, aluminium, …) are incorporated through the solution doping technique [xvii]. The last core layer is deposited at lower temperature than the preceding cladding layers, so that they are not fully sintered and left porous. Then the substrate tube is filled with an alcoholic solution of salts and allowed to impregnate the porous layers. After 1-2 hours, the solution is removed, the porous layer is dried and sintered. When the deposition is complete, the tube is collapsed at 2000°C into a preform. In our case, the typical length of the preform is about 30 cm and the diameter is 10 mm. The preform is then put into a furnace for drawing into fiber. The preform tip is heated to about 2000°C. As the glass softens, a thin drop falls by gravity and pulls a thin glass fiber. The diameter of the fiber is adjusted by varying the capstan speed. A UV-curable polymer is used to coat the fiber.

### Material characterizations

Refractive index profiles (RIP) of the preforms were measured using a York Technology refractive index profiler (P101), while the RIPs of the optical fibers were determined using a York Technology refractive index profiler (S14). The oxide core compositions of the samples were deduced from measurement of the RIP in the preform, knowing the correspondence between index rising and $AlO_{3/2}$, $GeO_2$, $PO_{5/2}$ concentration in silica glass from the literature [xviii,xix]. The composition was also directly measured on some preforms using electron probe microanalysis technique in order to compare results. A good agreement was found. The concentration of these elements is generally around few mol%. Luminescent ions concentrations are too low to be measured through the RIP. They were measured through absorption spectra. For example, $Tm^{3+}$ ion concentration has been deduced from the 785 nm ($^3H_6=>^3H_4$) absorption peak measured in fibers and using absorption cross-section reported in [xx]: $\sigma_{abs}$(785 nm) = $8.7 \times 10^{-25}$ m$^2$.

### Energy transfers in $Er^{3+}$ and $Yb^{3+}$:$Er^{3+}$ heavily doped silica fibers

The non-radiative energy transfer processes are well-known phenomena that influence the optical properties of doped-materials. The first theoretical basis appears in the 50's with the Förster-Dexter's model [xxi,xxii] that treats this process as the result of dipole-dipole and multipole-multipole interactions. Two energy transfer processes are described in Fig. 1. When pumping a co-doped material some ions are promoted in one of their excited level. If some ions are close to each other, their wave-functions interpenetrate and the energy stored in the excited level of the donor ions is non-radiatively transferred to a resonant level of the acceptor ions. This process was turned to good account in $Yb^{3+}$:$Er^{3+}$ co-doped silica fibers for high power fiber amplifier [xxiii] and laser [xxiv] applications : it takes advantage of the strong absorption cross-section of $Yb^{3+}$ at 980 nm and of the high efficiency of the energy transfer. In the case of high doping levels for both species, another non-radiative energy transfer process can take place and allows exciting a higher level of the acceptor ion : it is the double energy transfer process (DET) described by Auzel [xxv]. This process was first used to convert infrared light from LED to visible emission or to detect weak infrared signals with photomultipliers [xxvi,xxvii].

## Double energy transfer in $Er^{3+}$-doped fibers

The clustering effect in $Er^{3+}$-doped silica fibers is now a well-accepted phenomenon, and its detrimental influence on the 1550-nm gain transition of such fibers is well established [xxviii]. For simplicity, modeling of clusters has consisted of considering that a fraction of the dopants were organized in ion pairs [xxix], in which an immediate energy transfer leads to an instantaneous relaxation of one excited ion. This model is in very good agreement with the experimental results obtained for saturable absorption and for gain measurements at low $Er^{3+}$ doping levels as in fiber amplifiers. At higher doping levels, Ainslie *et al.* [xxx] showed that, in addition to the ions dispersed in the host, regions in which concentrations of rare-earth exceeding 40 wt% - called clusters – appear : in such a material the ion-pair model cannot be applied. We have developed a cluster model [xxxi] that differs from the ion-pair one by the fact that we consider that each ion of a cluster can efficiently transfer its energy to any of the other ions of the same cluster. When *n* ions of a cluster are excited, a succession of (*n*-1) fast relaxations by energy transfer leads to a situation in which all the ions of a cluster but one are de-excited. This model permits the determination of the proportion of the dopants organized in clusters and the transfer rate. In order to validate the model we realized a pump-absorption-versus-pump-power experiment with two fiber samples, *Er-1* and *Er-2*, doped with 100 ppm and 2,500 ppm of $Er^{3+}$, respectively (Fig. 2). This shows that the non-saturable absorption (NSA) grows dramatically with the $Er^{3+}$ concentration. We have attributed this behaviour to the presence of clusters containing a significant percentage of the dopants and in which efficient energy transfers allow these ions to relax rapidly after the absorption of a first pump photon.

**Double energy transfer in highly $Yb^{3+}$:$Er^{3+}$-co-doped fibers**

The green fluorescence of $Er^{3+}$-doped optical fiber is a well-known phenomenon in 800 nm-pumped erbium-doped fibers. This emission results from the excited state absorption phenomenon and is characteristic of the emission from the $^2H_{11/2}$ and $^4S_{3/2}$-levels and consequently can be observed with any pumping scheme leading to the population of these levels. We have studied how these levels can be excited by DET [xxxii] and a schematic energy diagram is shown in Fig. 3.

At low rare-earth concentrations, the large inter-ionic distances permit efficient single energy transfers, but the second energy transfer is very inefficient. For applications in which the green fluorescence is desirable, this second energy transfer must be enhanced. For that the rare-earth concentration must be as high as possible to reduce the distance between neighboring ions. In this case, a second phase, referred to as clusters, can appear in which the rare-earth ions concentration is particularly high. In order to quantify the fraction of active ions into clusters, we have studied the $Yb^{3+}$ and $Er^{3+}$ fluorescence dynamics in a highly co-doped fiber ([Yb]=[Er]=2,500 ppm) : the 1040 nm-fluorescence decay represents the population decay from the $^2F_{5/2}$-metastable level of $Yb^{3+}$, and that of the green-fluorescence represents the evolution of the $^2H_{11/2}$ and $^4S_{3/2}$-populations of $Er^{3+}$. Our setup allows simultaneous measurements of the counter-propagative visible emission and the lateral infrared emission. The experimental curves show two typical decays. Fitted with our rate equations model [xxxii], they revealed that roughly 50% of both ions are organized in clusters in the co-doped fiber. This high percentage must be associated with very high Yb-Er transfer rates ($3 \times 10^6$ s$^{-1}$), one order of magnitude superior to the $Er^{3+}$:$^4I_{11/2}$ intermediate level relaxation rate ($3.7 \times 10^5$ s$^{-1}$): $Er^{3+}$ ions placed in their short lived $^4I_{11/2}$ state have a higher probability to be excited to the $^4F_{7/2}$ upper state than to relax spontaneously. The strong percentage of ions organized in clusters and the very high transfer rates are at the origin of the very good up-conversion efficiency.

**Thermalization effects between excited levels in doped fibers: temperature sensor based on fluorescence of Er$^{3+}$**

Though the rare-earth ions are never in thermodynamical equilibrium because of the metastability of some levels, it has been demonstrated that the populations of the $^2H_{11/2}$ and $^4S_{3/2}$ levels responsible for the green emission in Er$^{3+}$-doped fibers are in quasi-thermal equilibrium. This effect has been observed for the first time in fluoride glass fibers [xxxiii] and can be attributed to the relatively long lifetime of these levels (400 µs) in that host. In silica, in spite of the two orders of magnitude shorter lifetime, a fast thermal coupling between both levels has been proposed [xxxiv] and confirmed experimentally [xxxv] (Fig. 4). Indeed these levels can be considered to be in quasi-thermal equilibrium, because of the small energy gap between them, about 800 cm$^{-1}$, compared to the high energy gap between them and the nearest lower level, about 3000 cm$^{-1}$. In this case, the lifetime of these levels is sufficient (1 µs) to allow populating the upper level from the lower one by phonon induced transitions. Therefore $R$, the ratio of the intensities coming from both levels, can be written as:

$$R = \frac{I(^2H_{11/2})}{I(^4S_{3/2})} = \frac{\nu(^2H_{11/2})}{\nu(^4S_{3/2})} \frac{\sigma_e(^2H_{11/2})}{\sigma_e(^4S_{3/2})} \exp[-\Delta E / kT] \quad (1)$$

where $\nu$ is the frequency, $\sigma_e$ the emission cross section, $k$ the Boltzmann constant, $\Delta E$ the energy gap between the two levels and $T$ the temperature in degrees Kelvin. In Fig. 4 we show that the experimental data can be fitted by a function in agreement with Equation (1). This is another example of an energy transfer process, this one being assisted by phonons.

In order to take advantage of the high efficiency of the DET in highly co-doped Yb-Er doped fiber and of the thermalisation effect between the higher levels involved in the green fluorescence in this kind of fiber, we have developed a new temperature sensor, unsensitive for strain. The dynamic obtained was 11 dB in Fig. 4 over the shown temperature range, leading to a mean rate of change of the green intensity ratio of approximately 0.016 dB/K at 300 K. Several temperature cycles have been carried out and we have observed a good repeatability. As for the stability, no modifications have been observed on the two intensities when the fiber was heated during several hours at temperatures up to 600°C. Due to the strong absorption of the doped fiber in the signal wavelength range - the green emissions corresponding to transitions downto the fundamental level - and to the 15 dB/km intrinsic absorption of the transparent fiber in the same wavelength range, such a device would be limited to a point sensor.

We have developed a new sensor based on the 1.13 µm and 1.24 µm emission lines, coming from the same levels [xxxvi]. These lines present the same temperature behaviour as the green ones. As the lower level of these transitions is the $^4I_{11/2}$-level and not the fundamental one (Fig. 3), the signals are absorption free and their wavelengths correspond to a transparency region of the intermediate fibers. These arguments have permitted the development of an efficient quasi-distributed configuration without limitation on the sensing line length : the short lifetime of upper levels (1 µs) could allow realizing a sensors network. Each sensitive head is separated from its neighbors by a 100-meter long transparent silica fiber in order to time-resolve the counter-propagative signals.

**Conclusion**

Energy transfer processes in rare-earth-doped materials have been studied since the middle of the 20$^{th}$ century. At the beginning, the applications of DET were mainly conversion of infrared light to visible emission or detection of weak infrared signals with photomultipliers. A renewal interest appears with the development of optical fibers in which high power density can be achieved: single energy transfer allows improvement of high power fiber amplifiers and laser and DET permits realizing point and quasi-distributed fiber sensors.

**Local structure, valency states and spectroscopy of transition metal ions**

Optical fiber materials with very broad-band gain are of great interest for many applications. Tunability in RE-doped fiber devices is already well established, but limited by shielding of the optically active electronic orbitals of RE ions. Optically active, unshielded orbitals are found in transition metal (TM) ions. Some TM-doped bulk solid-state lasers materials, such as $Cr^{4+}$:YAG, have demonstrated very good results as broad-band gain media [xxxvii]. Tentatives with other TM ions, like $Ni^{2+}$ in vitroceramics fiber are also promising [xiv]. More recently, a 400-nm emission bandwidth was observed from a fiber whose Cr-doped core was made of $Y_2O_3$:$A_2O_3$:$SiO_2$ obtained by a rod-in-tube technique using a $Cr^{4+}$:YAG rod as core material and a silica tube as cladding material [xxxviii].

Little literature exists on chromium- and other TM-doped vitreous bulk silica, although this issue was addressed in the 70's [xxxix] to improve transmission of silica optical fibers. Some reports on chromium-doped glasses have already shown evidence of absorption and near-infrared (NIR) fluorescence due to $Cr^{4+}$ in these materials [xl,xli]. However their compositions and preparation techniques greatly differ from those of silica optical fibers. Therefore, some basic studies on the optical properties of TM ions in silica-based optical fibers are needed. In particular, the final TM oxidation state(s) in the fiber core strongly depend(s) on the preparation process. Also, the optical properties (absorption and luminescence) of one particular oxidation state of a TM ion varies from one host composition and structure to another, due to variations of the crystal-field (so-called ligand field in glass) [xlii]. Hence the interpretation of absorption and emission spectroscopy is difficult. Because no luminescence spectroscopy of the TM-doped silica fibers had been reported before, we have contributed to explore this field. We have studied the influence of the chemical composition of the doped region on the Cr-oxidation states and on the spectroscopic properties of the samples. We have also studied the optical properties versus the experimental conditions (temperature and pump wavelength). We describe the experimental details specifically used for TM-doped fibers, then we summarize all results and interpretations.

**Fabrication and characterization of Chromium-doped samples**

The preforms and fibers were prepared as described in §0 Chapitre:, using $Cr^{3+}$-salt alcoholic doping solution and oxygen or nitrogen (neutral) atmosphere for the drying-to-collapse stages. Three different types of samples containing Ge or/and Al were prepared, referred to as *Cr(Ge)*, *Cr(Ge-Al)* and *Cr(Al)*, respectively. The total chromium concentration (*[Cr]*) was varied from below 50 mol-ppm to several thousands mol ppm. Above several $100^s$ mol ppm, preform samples had evidence of phase separation causing high background optical losses, whereas fibers (few 10s mol ppm) did not show phase separation and had low background losses (<1 dB/m). The oxidation states of Cr and their relative concentrations were analyzed by Electron Paramagnetic Resonance, whereas the absolute content of all elements (including Cr) was analyzed by Plasma Emission Spectroscopy.

Absorption spectra were analyzed using the Tanabe-Sugano (T.-S.) formalism [xliii] to compare our assignments to optical transitions with reports on $Cr^{3+}$- and $Cr^{4+}$-doped materials. This formalism helps predicting the energy of electronic states of a TM in a known ligand field symmetry as a function of the field strength *Dq* and the phenomenologic *B* and *C* so-called Racah parameters (all in $cm^{-1}$, Fig. 5). The *Dq/B* ratio allows the qualitative determination of some optical properties of TM ions, such as strength, energy and bandwidth of optical transitions. We have also estimated the absorption cross-sections using results from composition and valency measurements. Absorption and emission spectroscopies including decay measurements were performed on both preforms and fibers, at room temperature (RT) and low temperature (LT, either 12 or 77 K), using various pump wavelengths: 673, 900 and 1064 nm. Full details of the experimental procedures are given in [xliv,xlv,xlvi].

**Principal results**

By slightly modifying the concentration in germanium and/or aluminium in the core of the samples, their optical properties are greatly modified. In particular, we have shown that:

i) Only $Cr^{3+}$ and $Cr^{4+}$ oxidation states are stabilized. $Cr^{3+}$ is favoured by Ge co-doping, and lies in octahedral site symmetry (*O*), as in other oxide glasses [xlvii]. $Cr^{4+}$ is present in all samples. This valency is promoted by Al co-doping or when *[Cr]* is high, and lies in a distorted tetrahedral site symmetry ($C_s$) [xlviii,xlix]. The low-doped *Cr(Al)* samples contain only $Cr^{4+}$ and their absorption spectra are similar to those of aluminate [xl] and alumino-silicate glasses [xli] and even crystalline YAG ($Y_3Al_5O_{12}$) [l]. Glass modifiers like Al induce major spectroscopic changes, even at low concentrations (~1-2 mol%). This would help engineering the Chromium optical properties in silica-based fibers, using possibly alternative modifiers.

ii) The absorption spectra have been interpreted and optical transitions assigned for each present valency state (Fig. 6). The absorption cross sections curves ($\sigma_{abs}$) were estimated. For $Cr^{3+}$, $\sigma_{abs}(Cr^{3+}, 670\text{ nm}) = 43 \times 10^{-24}\text{ m}^2$ is consistent with reported values in other materials, such as ruby [li] and silica glass [xxxix], while $\sigma_{abs}(Cr^{4+}, 1000\text{ nm}) \sim 3.5 \times 10^{-24}\text{ m}^2$ is lower than in reference crystals for lasers [lii] or saturable absorbers [liii], but consistent with estimated values in alumino-silicate glass [xli].

iii) Using the T.-S. formalism, we found $Dq/B = 1.43$ which is lower than the value were $^3T_2$ and $^1E$ levels cross ($Dq/B = 1.6$, Fig. 5). As a consequence, the expected emission is along the $^3T_2 \rightarrow {}^3A_2$ transition as a broad featureless NIR band. No narrow emission line from the $^1E$ state is expected, in agreement with fluorescence measurements. $Dq/B$ is lower than those reported for $Cr^{4+}$ in laser materials like YAG and Forsterite [xlviii].

iv) The LT fluorescence from $Cr^{4+}$ spreads over a broad spectral domain, from 850 to 1700 nm, and strongly varies depending on core chemical composition, *[Cr]* and $\lambda_p$ (pump wavelength). The observed bands were all attributed to $Cr^{4+}$ ions, in various sites. Fig. 7 shows the fluorescence spectra of $Cr^{4+}$ in two different types of samples and in various experimental conditions. Possible emission from other centers ($Cr^{3+}$, $Cr^{5+}$, $Cr^{6+}$) was discussed, but rejected [xlvi]. The fluorescence sensitivity to *[Cr]* and $\lambda_p$ suggests that Cr-ions are located in various host sites, and that several sites are simultaneously selected by an adequate choice of $\lambda_p$ (like in *Cr(Ge-Al)*). It is also suggested that although Al promotes $Cr^{4+}$ over $Cr^{3+}$ when *[Cr]* is low, $Cr^{4+}$ is also promoted in Ge-modified fibers at high *[Cr]*.

v) The strong decrease of fluorescence from LT to RT is attributed to temperature quenching caused by multiphonon relaxations, like in crystalline materials where the emission drops by typically an order of magnitude from 77 K to 293 K [xxxvii].

vi) The LT fluorescence decays are non-exponential (Fig. 8) and depend on *[Cr]* and $\lambda_s$. The fast decay part is assigned to Cr clusters or $Cr^{4+}$-rich phases within the glass. The 1/e-lifetimes ($\tau$) at $\lambda_s = 1100$ nm are all within the 15-35 µs range in Al-containing samples, whereas $\tau \sim$ 3-11 µs in *Cr(Ge)* samples, depending on *[Cr]*. The lifetime of isolated ions ($\tau_{iso}$), measured on the exponential tail decay curves (not shown) reach high values: $\tau_{iso} \sim$ 200 to 300 µs at $\lambda_s \sim$1100 nm, $\tau_{iso} \sim$70 µs at $\lambda_s \sim$1400 nm. In the heavily-doped *Cr(Ge)* samples, $\tau_{iso}$ is an order of magnitude less. Hence, $Cr^{4+}$ ions are hosted in various sites: the lowest energy ones suffer more non-radiative relaxations than the higher energy ones. Also presence of Al improves the lifetime, even at high *[Cr]*. It is estimated that at RT, lifetime $\tau$ would be of the order of 1 µs or less. This fast relaxation time, compared to RE ions (~1 ms) has been implemented as a fiberized saturable absorber in a passively Q-switched all-fiber laser [liv].

**Conclusion**

The observed LT fluorescence of $Cr^{4+}$ is extremely sensitive to glass composition, total Cr concentration and excitation wavelength. Using Aluminum as a glass network modifier has advantages: longer excited state lifetime and broader fluorescence bandwidth than in Germanium-modified silica. A combination of Al and Ge glass modification induces the broadest fluorescence emission in the NIR range, to our knowledge, exhibiting a 550 nm-bandwidth. However, increasing the quantum efficiency is now necessary for practical fiber amplifiers and light sources. Further investigations concluded to the necessity of local surrounding TM ions with a different material, i.e. having sensibly different chemical and physical properties compared to pure silica, in order to improve the local site symmetry and hence minimize NRD. Preliminary implementation of this principle was reported recently, concerning $Cr^{3+}$ ions in post-heat-treated Ga-modified silica fibers [lv]. When engineering of the local dopant environment will be possible, then practical TM-doped silica-based amplifying devices will be at hand.

**Phonon interactions / non-radiative relaxations: improvement of $Tm^{3+}$ efficiency**

Thulium-doped fibers have been widely studied in the past few years. Because of $Tm^{3+}$ ion rich energy diagram, lasing action and amplification at multiple infrared and visible wavelengths are allowed. Thanks to the possible stimulated emission peaking at 1.47 µm ($^3H_4$ => $^3F_4$, see Fig. 9), discovered by Antipenko et al. [lvi], one of the most exciting possibilities of $Tm^{3+}$ ion is amplifying optical signal in the S-band (1.47‑1.52 µm), in order to increase the available bandwidth for future optical communications. Unfortunately, the upper $^3H_4$ level of this transition is very close to the next lower $^3H_5$ level so non-radiative de-excitations (NRD) are likely to happen in high phonon energy glass host, causing detrimental gain quenching.

**Oxide modifiers influence on the $^3H_4$-level lifetime**

To address this problem, we have studied the effect of some modifications of $Tm^{3+}$ ion local environment. Keeping the overall fiber composition as close as possible to that of a standard silica fiber, we expect to control the rare-earth spectroscopic properties by co-doping with selected modifying oxides. We have studied the incorporation of modifying elements compatible with MCVD. $GeO_2$ and $AlO_{3/2}$ are standard refractive index raisers in silica. $AlO_{3/2}$ is also known to improve some spectroscopic properties of $Er^{3+}$ ion for C-band amplification [i] and to reduce quenching effect through clustering in highly rare-earth-doped silica [lvii]. Both oxides have a lower maximum phonon energy than silica. We use high phonon energy $PO_{5/2}$ as opposite demonstration. $GeO_2$ and $PO_{5/2}$ concentrations are 20 and 8 mol%, respectively. $AlO_{3/2}$ concentration is varied from 5.6 to 17.4 mol%. $Tm^{3+}$ concentration is less than 200 mol ppm.

To investigate the role of the modification of the local environment, decay curves of the 810 nm fluorescence from the $^3H_4$ level were recorded. All decay curves measured are non-exponential. This can be attributed to several phenomena and will be discussed in this article. Here, we study the variations of 1/e lifetimes ($\tau$) versus concentration of oxides of network modifiers (Al or P) and formers (Ge). The lifetime strongly changes with the composition of the glass host. The most striking results are observed within the *Tm(Al)* sample series: $\tau$ linearly increases with increasing $AlO_{3/2}$ content, from 14 µs in pure silica to 50 µs in sample *Tm(Al)* containing 17.4 mol% of $AlO_{3/2}$. The lifetime was increased about 3.6 times. The lifetime of the 20 mol% $GeO_2$ doped fiber *Tm(Ge)* was increased up to 28 µs whereas that of the 8 mol% $PO_{5/2}$ doped fiber *Tm(P)* was reduced down to 9 µs. We see that aluminum codoping seems the most interesting route among the three tested codopants.

**Non-exponential shape of the 810-nm emission decay curves**

All fluorescence decay curves from the $^3H_4$ level are non-exponential. We have investigated the reasons for this non-exponential shape in various silica glass compositions. We observed that the decay curve shape depends only on the Al-concentration, even in the presence of Ge or P in samples *Tm(Ge)* and *Tm(P)*, respectively [lviii]. It is thought that $Tm^{3+}$ ions are inserted in a glass which is characterized by a multitude of different sites available for the rare-earth ion, leading to a multitude of decay constants. This phenomenological model was first proposed by Grinberg *et al.* and applied to $Cr^{3+}$ in glasses [lix]. Here we apply this model, for the first time to our knowledge, to $Tm^{3+}$-doped glass fibers. In this method, a continuous distribution of lifetime rather than a number of discrete contributions is used. The advantage of this method is that no luminescence decay model or physical model of the material is required *a priori*. The luminescence decay is given by:

$$I(t) \approx \sum_i A_i \exp[-t/\tau_i] \qquad (2)$$

where $A(\tau)$ is the continuous distribution of decay constant.

The procedure for calculating $A(\tau)$ and the fitting algorithm are described in detail in [lix]. For the fitting procedure, we considered 125 different values for $\tau_i$, logarithmically spaced from 1 to 1000 µs. By applying this procedure to all the decay curves, a good matching was generally obtained. For a given composition (Fig. 10), we can notice two main distributions of the decay constant. With the aluminium concentration, they increase from 6 to 15 µs and from 20 to 50 µs, respectively. For the highest aluminium concentration (9 mol%, in *Tm(Ge)* and

*Tm(P)*), these two bands are still present (not shown in the figure). One is around 10 µs and the second one spreads from 30 to 100 µs, for both compositions (*Tm(Ge)* and *Tm(P)*). According to the phenomenological model, the width of the decay constants distribution is related to the number of different sites. The large distribution around 80 µs is then due to a large number of sites available with different environments. It is however remarkable that this distribution at ≈80 µs is very similar in both sample types. From the $Tm^{3+}$ ion point of view (considering luminescence kinetics), *Tm(Ge)* and *Tm(P)* glasses seem to offer the same sites.

The meaning of the decay constant values is now discussed. Lifetime constants obtained from the fitting can be correlated with the one expected for Thulium located in a pure silica or pure $Al_2O_3$ environment. The $^3H_4$ lifetime is calculating by using this equation:

$$1/\tau = 1/\tau_{rad} + W_{nr} \qquad (3)$$

where $\tau_{rad}$ corresponds to the radiative lifetime which is given to be 670 µs in silica [lx]. $W_{nr}$ is the non-radiative decay rate, expressed as [lxi]:

$$W_{nr} = W_0 \times \exp[-\alpha(\Delta E - 2E_p)] \qquad (4)$$

where $W_0$ and $\alpha$ are constants depending on the material, $\Delta E$ is the energy difference between the $^3H_4$ and $^3H_5$ levels and $E_p$ is the phonon energy of the glass. $W_0$ and $\alpha$ were estimated for different oxide glasses [lxi, lxii]. The energy difference $\Delta E$ was estimated by measuring the absorption spectrum of the fibers. When Al concentration varies, this value is almost constant around 3700 $cm^{-1}$ [lxiii].

With these considerations, the $^3H_4$ expected lifetime can be calculated. In the case of silica glass, $\tau_{silica}$ = 6 µs and for an $Al_2O_3$ environment, $\tau_{alumina}$ = 110 µs. These two values are in accordance with the ones we obtained from the fitting procedure. The distribution of decay constant around 10 µs corresponds to $Tm^{3+}$ ions located in almost pure silica environment while the second distribution is attributed to $Tm^{3+}$ located in $Al_2O_3$-rich sites.

**Conclusion**

By adding oxide network modifiers or formers, we demonstrated that aluminium is the most efficient to improve the $^3H_4$ level-lifetime. This was attributed to a lower local phonon energy. Potential of the amplification in the S-band was then investigated. In the fiber with the highest aluminium concentration, gain curve was measured. Although excitation wavelength (1060 nm), refractive index profile and Thulium concentration were not optimized, a gain of 0.9 dB was obtained at 1500 nm [lxiv]. With a numerical model of the TDFA that we developed [lxv], we estimated that a gain higher than 20 dB is reachable in a silica-based TDFA.

**Rare-earth-doped dielectric nanoparticles**

Erbium-doped materials are of great interest in optical telecommunications due to the $Er^{3+}$ intra-4f emission at 1.54 µm. Erbium-Doped Fiber Amplifiers (EDFA) were developed in silica glass because of the low losses at this wavelength and the reliability of this glass. Developments of new rare-earth doped fiber amplifiers aim to control their spectroscopic properties: shape and width of the gain curve, optical quantum efficiency, .... Standard silica glass modifiers, such as aluminum, give very good properties to available EDFA. However, for more drastic spectroscopic changes, more important modifications of the rare-earth ions local environment are required. To this aim, we present a fiber fabrication route creating rare-earth doped calco-silicate or calco-phospho-silicate nanoparticles (NP) embedded in silica glass.

**Nanostructured fibers preparation**

In the chosen route, NP are not prepared *ex-situ* and incorporated into the perform. To prepare them, we take advantage of the heat treatement occurring during the MCVD process. Their formation is based on the basic principle of phase separation. On the basis of thermodynamical data such as activity coefficient, entropy of mixing, enthalpy of mixing and Gibbs-free energy change, the phase diagram of the $SiO_2$-CaO binary compound was derived using Factstage software (Fig. 11). A miscibility gap is found when the CaO concentration is between 2 and 30 mol%. In this region, CaO droplets are formed, like oil in water. Such phenomenon is expected during perform fabrication as temperature reaches 2000°C during collapsing passes.

For Calcium doping, $CaCl_2$ salt was added to the $Er^{3+}$ containing soaking solution. Four $CaCl_2$ concentrations were studied (0, 0.001, 0.1 and 1 mol/l). Ge and P were also added by MCVD. When the Ca concentration was increased in the doping solution, the aspect of the central core of the preform turned from transparent to milky. This variation is explained by the structural changes of the core. For preforms with calcium concentration higher than 0.01 mol/l, NP were observed by Transmission Electron Microscopy (TEM) on preform samples (Fig. 12). We can clearly observe polydisperse spherical NP with an estimated mean diameter of 50 nm. Smaller particles of 10 nm are visible. The size of the biggest particles was around 200 nm (not shown in Fig. 12). When the Ca concentration decreases, the size distribution of the particles is nearly identical but the density is lower. The composition of the core was investigated by Energy Dispersive X-ray analysis: the NP contained equal amounts of Ca, P and Si cations, whereas only Si cations was detected in the surrounding matrix. Ge seemed to be homogeneously distributed over the entire glass. The most important finding is that $Er^{3+}$ ions and Ca were detected only within the NP.

### Erbium emission characterizations

Spectroscopic characterizations on the emission line associated to the $^4I_{13/2}$-$^4I_{15/2}$ transition at 1.54 $\mu$m were made at room temperature on Er-doped samples with (sample *A*) and without (sample *B*) calcium. The results are shown in Fig. 13 where we evidence the fact that the emission spectrum of sample *A* is broader than that of sample *B*. To explain these differences we have studied the $Er^{3+}$ local environment. EXAFS measurements at the *Er-$L_{III}$* edge (E=8358 eV) were carried out at the GILDA-CRG beamline at the European Synchrotron Radiation Facility. In sample *A* the rare-earth is linked to O atoms in the first coordination shell and to Si or P atoms (these two atoms can not be distinguished due to the similar backscattering amplitude and phase) in the second shell in a way similar to that already observed in silicate glasses [lxvi], phosphate glasses [lxvii]. Also the structural parameters (about 7 O atoms at 2.26 Å and Si (or P) atoms at 3.6 Å corresponding to a Er-O-Si (or P) bond angle of ≈140 deg) are in good agreement with the cited literature. Si(or P) atoms are visible as they belong to the same $SiO_4$ (or $PO_4$) tetrahedron as the first shell O atoms but no further coordination shells are detected. This permits to state that an amorphous environment is realized around $Er^{3+}$ ions. On the other hand sample *B* presents a completely different EXAFS signal that is well comparable with the spectrum of $ErPO_4$. This means that Er in this case is inserted in a locally well ordered phase of about a few coordination shells (around 4-5 Å around the absorber). The fact that TEM on this sample reveals a uniform sample is not in contradiction with this result; it just means that this phase is not spatially extended to form nm-sized NP (in our TEM analyses the spatial resolution is limited to few nm) but the ordering is extremely local, i.e. it is limited to only a few shells around the rare-earth ion.

From these considerations, the broadening of the emission spectrum observed in sample *A* can be attributed to an inhomogeneous broadening due to $Er^{3+}$ ions located in a more disordered environment compare to sample *B*. Here we see that the cumulated effects of Ca and P within the Er-doped NP both amorphize the material structure around $Er^{3+}$ ions and increase the fluorescence inhomogeneous broadening.

**Conclusion**

In this paragraph, we have demonstrated that through the phase separation mechanism, nanoparticles can be obtained in preforms by adding Calcium. $Er^{3+}$ ions are found to be located only into these nanoparticles. An inhomogeneous broadening of the emission band is observed, associated to $Er^{3+}$ ions located in a more disordered environment compare to silica. This feature is particularly interesting in the production of materials for Wavelength Division Multiplexing applications, such as Erbium-Doped Fiber Amplifier with a broader band gain.

**Perspectives and conclusion**

The choice of a glass to develop new optical fiber component is most of the time a result of compromises. Silica glass is the most widely used for its many advantages (reliability, low cost fabrication, …). However, it suffers from different drawbacks, such as high phonon energy or low luminescent ions solubility, which affect quantum efficiency or emission bandwidth of luminescent ions, for example. We have shown in several cases that spectroscopic properties of dopants are not directly related to the average properties of the doped glass, but to their local environment. Indeed, by slightly modifying the silica composition, we succeeded to control the Chromium valence state and improve the Thulium emission efficiency. Moreover, we present interest of high doping level to take advantage of energy transfers. Then, nanostructuration of doped fiber is proposed as a new route to 'engineer' the local dopant environment. All these results will benefit to optical fiber components such as lasers, amplifiers and sensors, which can now be realized with silica glass.

**Figures captions**

Fig. 1: Schematic energy diagram of (a) single energy transfer between two ions, (b) double energy transfer.

Fig. 2 : Absorption of the *Er-1* and *Er-2* fibers vs launched pump power. Squares: experimental data; solid lines: cluster model for 0%, 10% and 52% of $Er^{3+}$-ions in clusters; vertical arrows: non-saturated absorption as a difference with simulation for 0% cluster.

Fig. 3 : Energy scheme for the DET process. Level energies are in $cm^{-1}$; their lifetimes between round brackets.

Fig. 4 : Natural logarithm of measured intensity ratio (R) plotted against the inverse of temperature.

Fig. 5: Normalized Tanabe-Sugano energy level diagram for $Cr^{4+}$ in tetrahedral ligand field ($T_d$ symmetry) showing the energy states of interest, for $C/B = 4.1$. The free ion states are shown on the left of the ordinate axis. The dashed line ($Dq/B = 1.43$) reveals the relative positions of the states found for $Cr^{4+}$ in the silica-based samples: the first excited state level is $^3T_2(^3F)$.

Fig. 6:. Background corrected absorption from (left) a *Cr(Ge)* preform ($[Cr] = 1400$ ppm) and (right) a *Cr(Al)* fiber ($[Cr] = 40$ ppm). Circles: experimental data; Solid lines: adjusted bands to $Cr^{3+}$ (left) and $Cr^{4+}$ (right) transitions, respectively; and resulting absorption spectra. Assignments are indicated from the ground level $Cr^{3+}:^4A_2$ or $Cr^{4+}:^3A_2$ to the indicated excited level, respectively. The $Cr^{4+}:^3T_2$ level three-fold splitting is due to distorsion from perfect tetrahedral symmetry. The spin-forbidden $Cr^{3+}:^4A_2 \rightarrow {^2E}$ and $Cr^{4+}:^3A_2 \rightarrow {^1E}$ transitions are not visible and overlapping with the intense spin-allowed transitions.

Fig. 7: Fluorescence spectra: (a) fiber *Cr(Al)*:*[Cr]* = 40 ppm, $\lambda_p$ = 900 nm, *T*=77 K, (b) preform *Cr(Ge-Al)*: *[Cr]* = 300 ppm, $\lambda_p$ = 673 nm, *T*=12K.

Fig. 8: Fluorescence decays from *Cr(Al)* samples, $\lambda_p$ = 673 nm, *T*=12 K: (a) $\lambda_s$~1100 nm and *[Cr]*=40 ppm, (b) $\lambda_s$~1100 nm, *[Cr]*=4000 ppm, and (c) $\lambda_s$~1400 nm, *[Cr]*=4000 ppm

Fig. 9: Schematic energy diagram of $Tm^{3+}$ ion, showing the relevant multiplets. Solid arrows: absorption and emission optical transitions; thick arrow: NRD (non-radiative de-excitation) across the energy gap between the $^3H_4$ and $^3H_5$ multiplets, $\Delta E$~ 3700 $cm^{-1}$.

Fig. 10 : Histograms of the recovered luminescence decay time distributions obtained for silica-based $Tm^{3+}$-doped fibers with phosphorus incorporated in the core and different $Al_2O_3$ concentration.

Fig. 11: Miscibility-gap in the derived phase-diagram of binary $SiO_2$-CaO glass

Fig. 12: TEM image from preform sample doped with Ca and P.

Fig. 13 : Room temperature emission spectra of Er-doped preform with (sample *A)* and without (sample *B)* Calcium. Samples were excited at 980 nm.

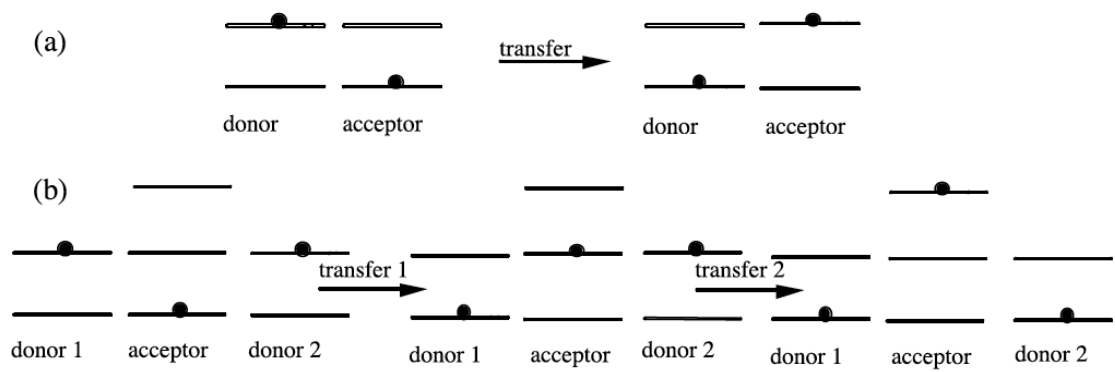

Fig. 1: Schematic energy diagram of (a) single energy transfer between two ions, (b) double energy transfer.

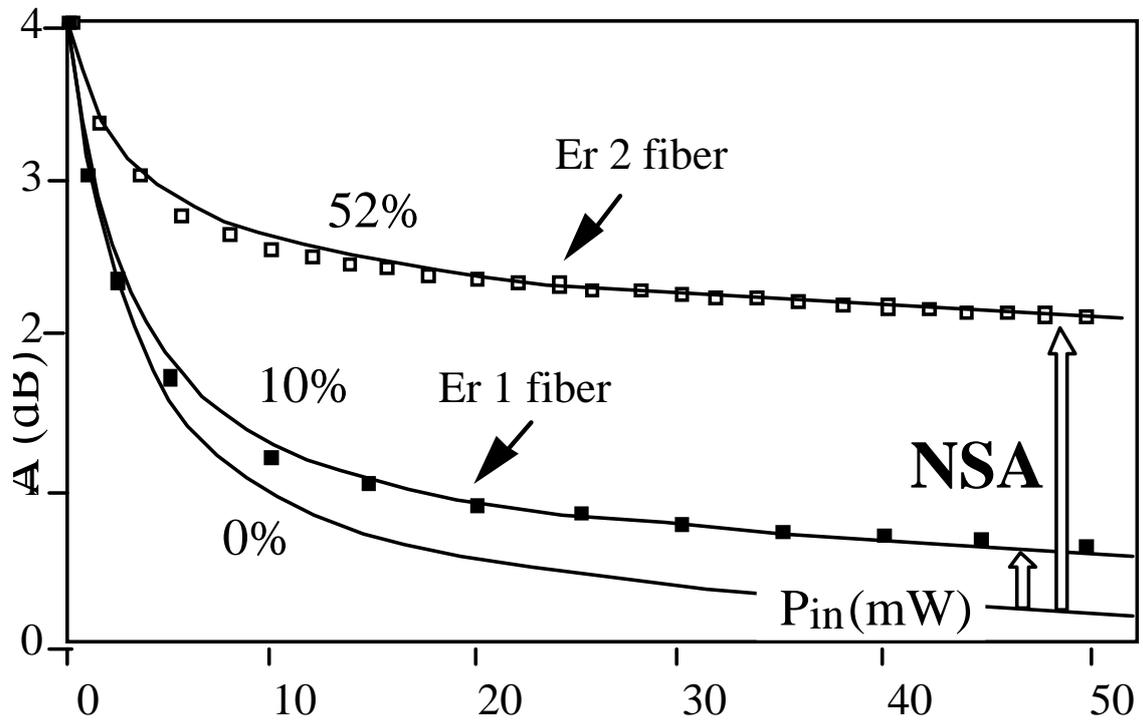

Fig. 2 : Absorption of the *Er-1* and *Er-2* fibers vs launched pump power. Squares: experimental data; solid lines: cluster model for 0%, 10% and 52% of $Er^{3+}$-ions in clusters; vertical arrows: non-saturated absorption as a difference with simulation for 0% cluster.

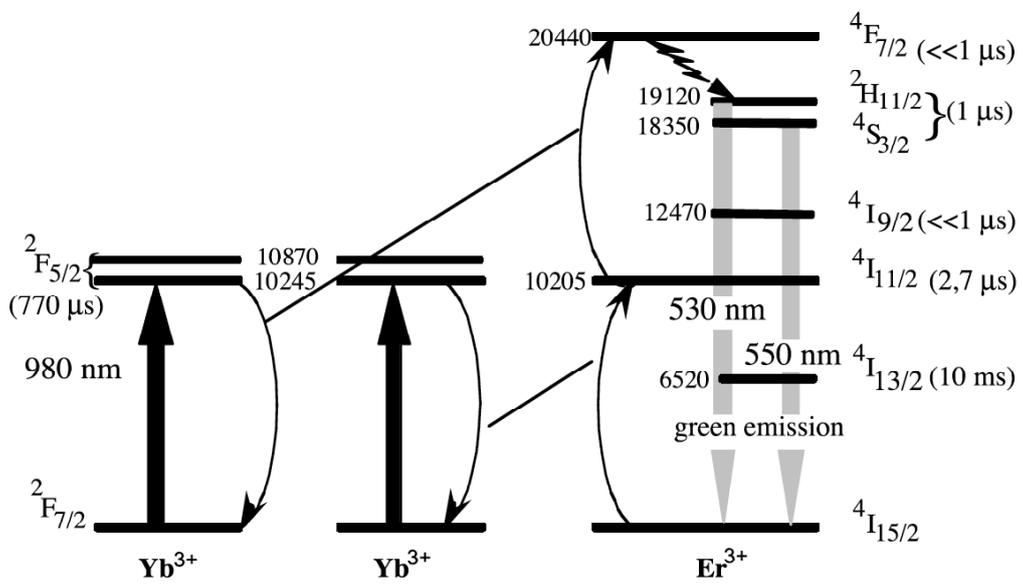

Fig. 3 : Energy scheme for the DET process. Level energies are in cm$^{-1}$; their lifetimes between round brackets.

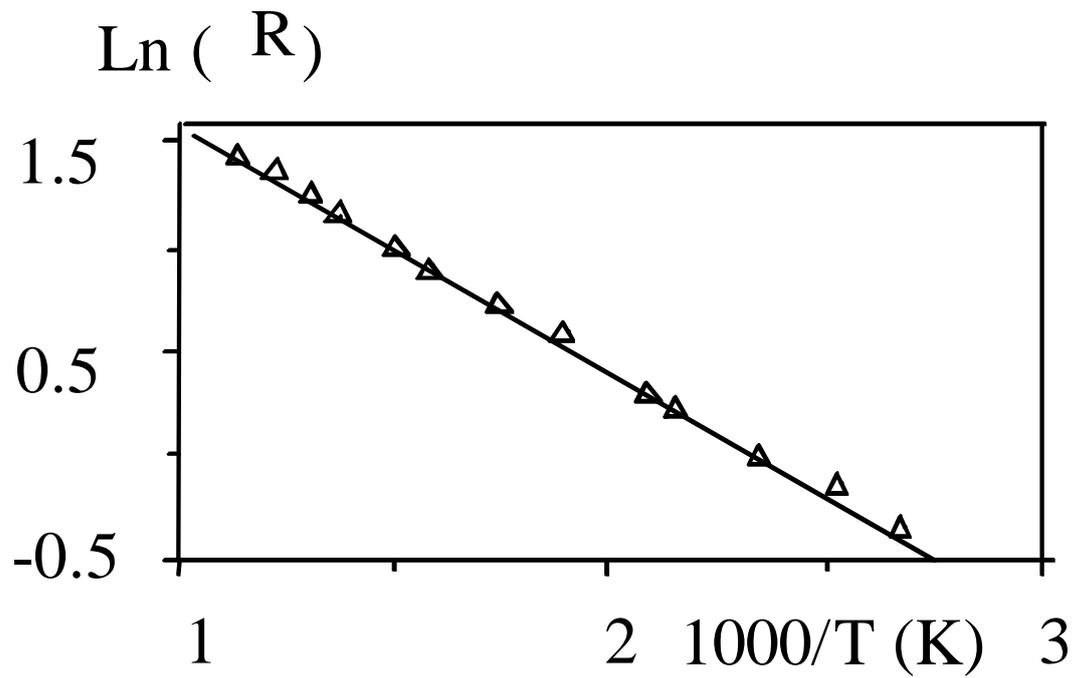

Fig. 4 : Natural logarithm of measured intensity ratio (*R*) plotted against the inverse of temperature.

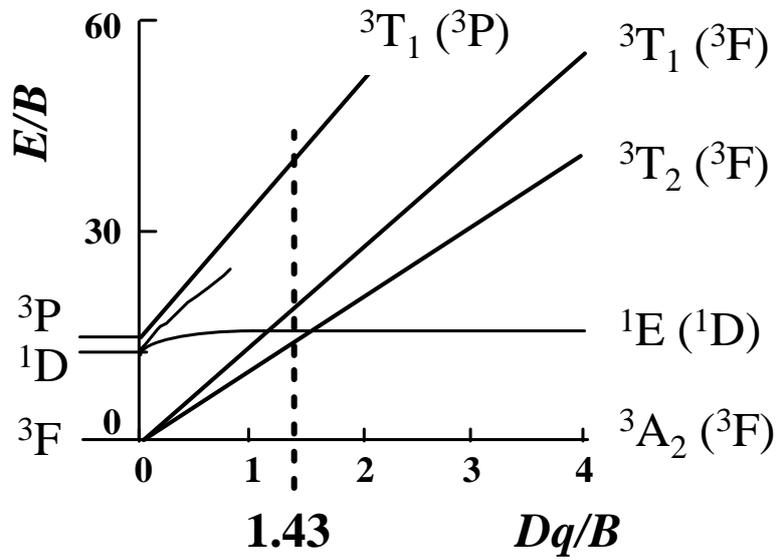

Fig. 5: Normalized Tanabe-Sugano energy level diagram for $Cr^{4+}$ in tetrahedral ligand field ($T_d$ symmetry) showing the energy states of interest, for $C/B = 4.1$. The free ion states are shown on the left of the ordinate axis. The dashed line ($Dq/B = 1.43$) reveals the relative positions of the states found for $Cr^{4+}$ in the silica-based samples: the first excited state level is $^3T_2(^3F)$.

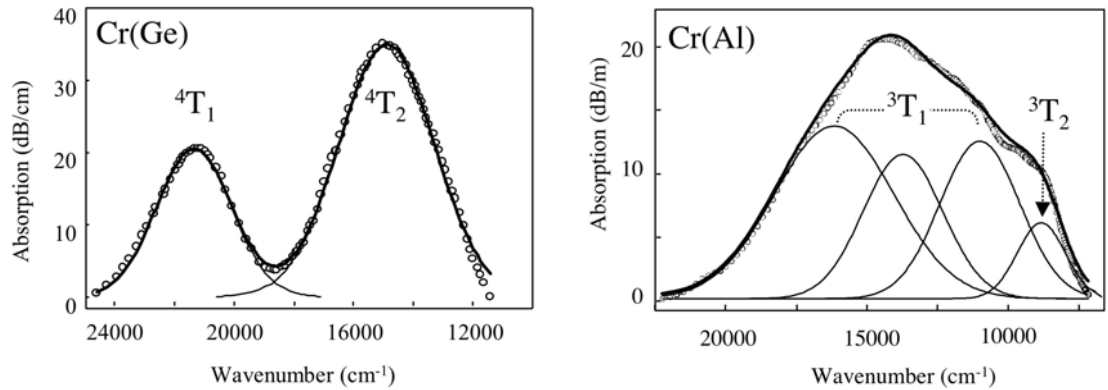

Fig. 6:. Background corrected absorption from (left) a *Cr(Ge)* preform (*[Cr]* = 1400 ppm) and (right) a *Cr(Al)* fiber (*[Cr]* = 40 ppm). Circles: experimental data; Solid lines: adjusted bands to $Cr^{3+}$ (left) and $Cr^{4+}$ (right) transitions, respectively; and resulting absorption spectra. Assignments are indicated from the ground level $Cr^{3+}:^4A_2$ or $Cr^{4+}:^3A_2$ to the indicated excited level, respectively. The $Cr^{4+}:^3T_2$ level three-fold splitting is due to distorsion from perfect tetrahedral symmetry. The spin-forbidden $Cr^{3+}:^4A_2 \rightarrow ^2E$ and $Cr^{4+}:^3A_2 \rightarrow ^1E$ transitions are not visible and overlapping with the intense spin-allowed transitions.

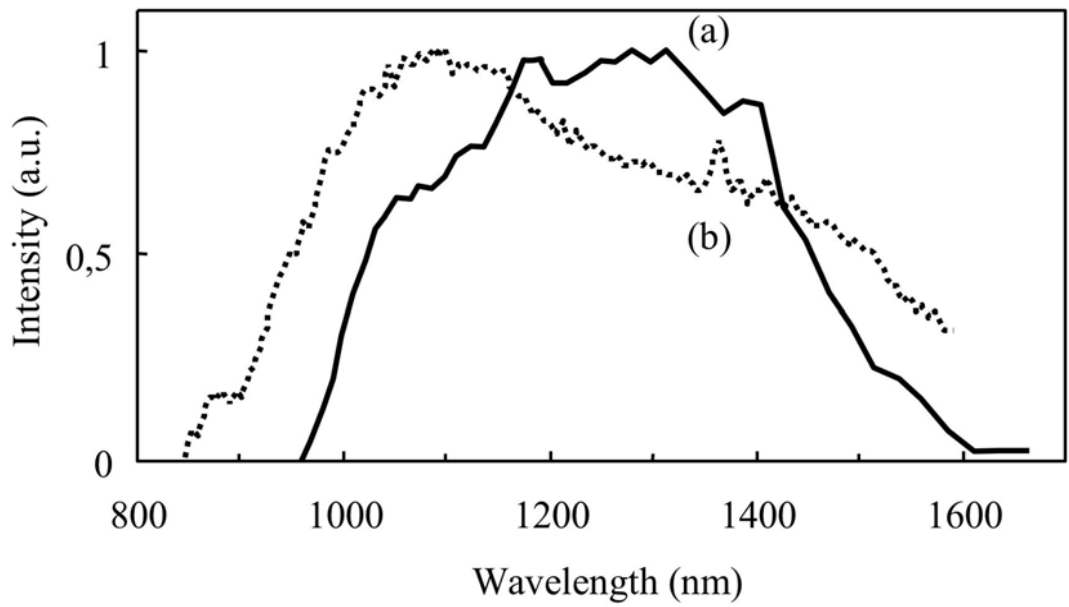

Fig. 7: Fluorescence spectra: (a) fiber *Cr(Al)*:*[Cr]* = 40 ppm, $\lambda_p$ = 900 nm, *T*=77 K, (b) preform *Cr(Ge-Al)*: *[Cr]* = 300 ppm, $\lambda_p$ = 673 nm, *T*=12K.

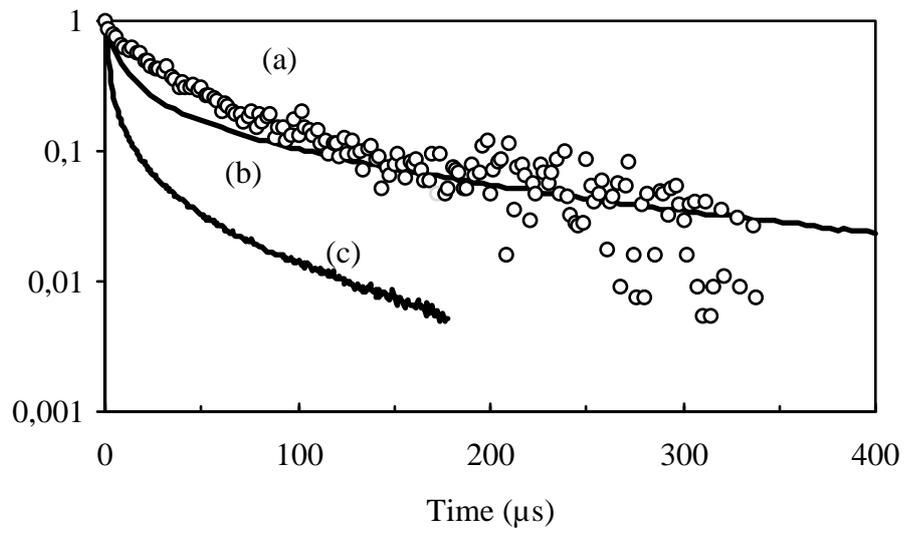

Fig. 8: Fluorescence decays from *Cr(Al)* samples, $\lambda_p = 673$ nm, $T=12$ K: (a) $\lambda_s \sim 1100$ nm and *[Cr]*=40 ppm, (b) $\lambda_s \sim 1100$ nm, *[Cr]*=4000 ppm, and (c) $\lambda_s \sim 1400$ nm, *[Cr]*=4000 ppm

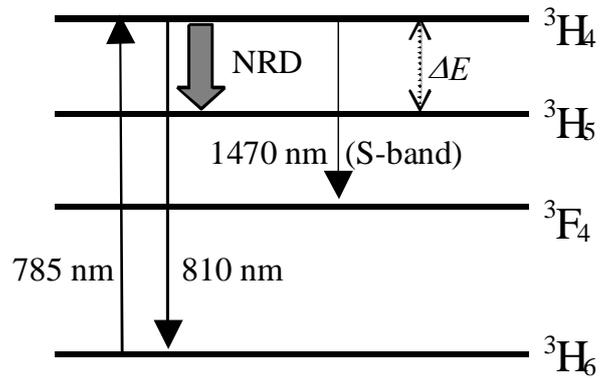

Fig. 9: Schematic energy diagram of Tm$^{3+}$ ion, showing the relevant multiplets. Solid arrows: absorption and emission optical transitions; thick arrow: NRD (non-radiative de-excitation) across the energy gap between the $^3H_4$ and $^3H_5$ multiplets, $\Delta E \sim 3700$ cm$^{-1}$.

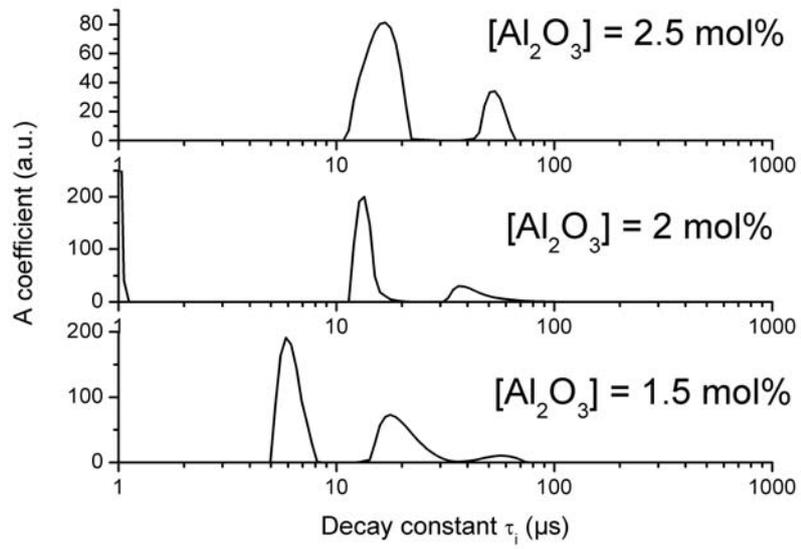

Fig. 10 : Histograms of the recovered luminescence decay time distributions obtained for silica-based $Tm^{3+}$-doped fibers with phosphorus incorporated in the core and different $Al_2O_3$ concentration.

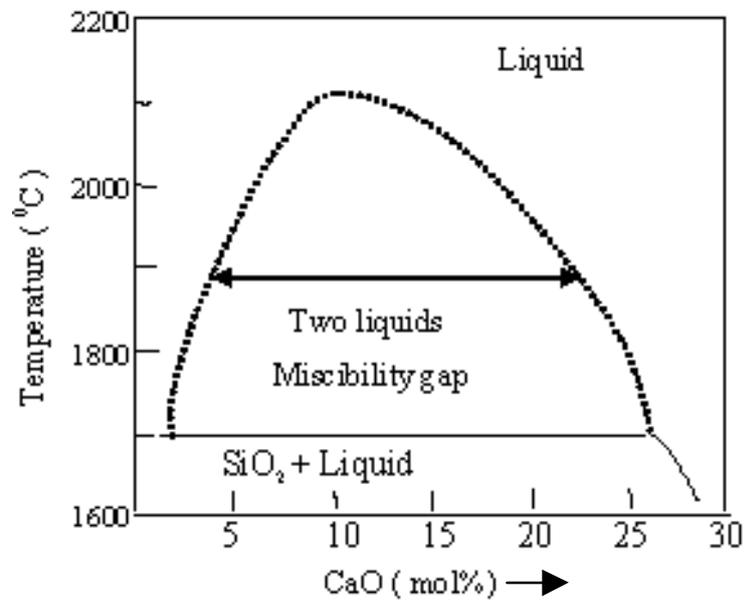

Fig. 11: Miscibility-gap in the derived phase-diagram of binary $SiO_2$-CaO glass

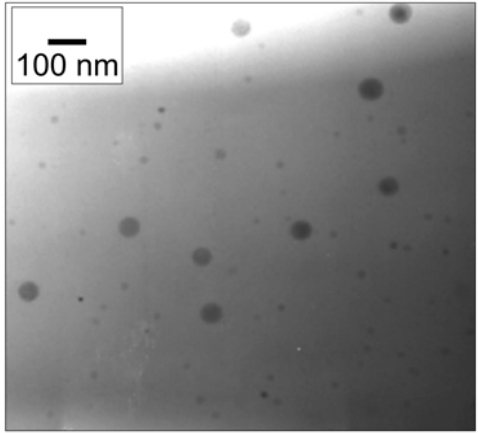

Fig. 12: TEM image from preform sample doped with Ca and P.

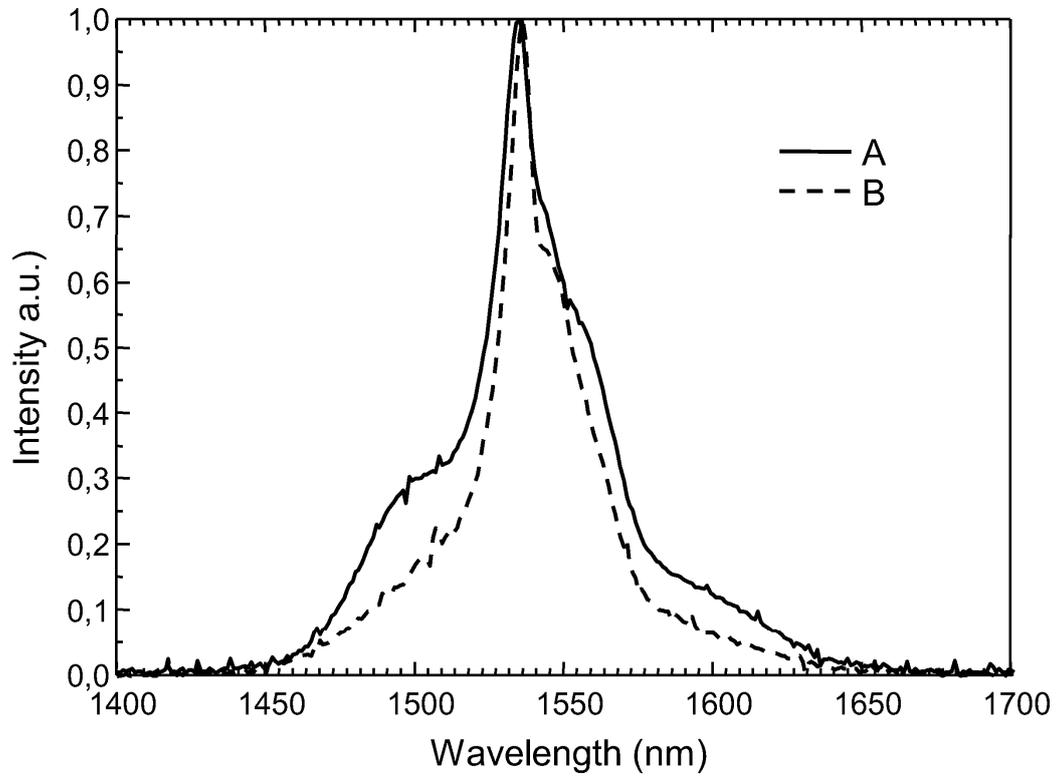

Fig. 13 : Room temperature emission spectra of Er-doped preform with (sample *A*) and without (sample *B*) Calcium. Samples were excited at 980 nm.